\documentclass[aps,prl,twocolumn,floatfix,longbibliography]{revtex4-1}
\usepackage{enumitem}
\usepackage{graphicx}
\usepackage{bm}
\usepackage{epsf}
\usepackage{ulem}
\usepackage[colorlinks = true]{hyperref}
\usepackage{amssymb}
\usepackage{amsmath}
\usepackage{graphicx}
\usepackage{rotating}
\usepackage{epsfig}
\usepackage{psfrag}
\usepackage{amsmath}
\usepackage{subfigure}
\newcommand{\bk}{{\bf k}}

\newcommand{\gtappr}{{{\lower4pt\hbox{$>$} } \atop \widetilde{ \ \ \ }}}

\DeclareGraphicsExtensions{.pdf}
\newlength{\figwidth}
\newcommand{\fg}[3]
{\begin{figure}[htb]\vspace*{-0cm}\centerline{\includegraphics[width=\figwidth]{#1}}\vskip
-0.2cm \caption{\label{#2}#3}\end{figure}}
\newcommand{\urs}{URu$_{2}$Si$_{2}$\ }

\begin{document}
\title{Thermodynamic Measurement of Angular Anisotropy at the Hidden Order Transition of \urs}
\author{Jennifer Trinh$^{1}$,  Ekkes Br{\"u}ck$^{2}$, Theo Siegrist$^{3,4}$, Rebecca Flint$^{5}$, Premala
Chandra$^{6}$, Piers Coleman$^{6,7}$ and Arthur P. Ramirez$^{1}$}
\affiliation{$^1$ Physics Department,  UC Santa Cruz, 
Santa Cruz, California, 95064, USA}
\affiliation{$^2$Fundamental Aspects of Materials and Energy, Faculty of Applied Sciences, TU Delft Mekelweg, 15, 2629 JB, Delft, The Netherlands}
\affiliation{$^3$National High Magnetic Field Laboratory, Florida State University, Tallahassee, Florida, 32310, USA \\ 
	$^4$Department of Chemistry and Biomedical Engineering, Florida State University, Tallahassee, Florida, 32310, USA}
\affiliation{$^5$Department of Physics and Astronomy,
	Iowa State University, Ames, Iowa, 50011}
\affiliation{$^6$Center for Materials Theory, Rutgers University, Piscataway, New Jersey, 08854, USA \\ 
$^7$Department of Physics, Royal Holloway, University of London, Egham, Surrey TW20 0EX, UK}
\date{\today}

\begin{abstract}
The heavy fermion compound URu$_{2}$Si$_{2}$ continues to attract great
interest due to the unidentified hidden order it develops
below 17.5K. The unique Ising character of the spin fluctuations and
low temperature quasiparticles is well established. We present detailed
measurements of the angular anisotropy of the nonlinear magnetization that
reveal a $\cos^{4}\theta$ Ising anisotropy both at and above the ordering
transition. With Landau theory, we show this implies a strongly Ising
character of the itinerant hidden order parameter.
\end{abstract}
\maketitle

Despite intensive theoretical and experimental efforts, the hidden 
order (HO) that develops below 17.5K in the heavy fermion superconductor
\urs remains unidentified thirty years after its original discovery
\cite{Mydosh:2011wm}.
The nature of the quasiparticle excitations and the broken symmetries 
associated with the HO phase are important questions for understanding not only HO but also the low-temperature exotic superconductivity.
While \urs is tetragonal above the HO, torque magnetometry \cite{Okazaki:2010tn}, 
cyclotron resonance\cite{Tonegawa:2012}, x-ray diffraction \cite{Tonegawa:2014}  and elastoresistivity 
measurements \cite{Riggs:jh} 
indicate fourfold symmetry-breaking in the basal plane.  However NMR and NQR studies suggest that this nematic signal decreases with 
increasing
sample size and also depends on sample quality, suggesting that the bulk is tetragonal \cite{Mito13,Kambe15}.

A number of measurements on \urs indicate Ising anisotropy, suggesting
that it is essential to understanding its HO. 
At the HO transition temperature $T_{c}$,  both the linear ($\chi_1$) 
and nonlinear
($\chi_3$) susceptibilities are anisotropic, with $\chi_3$ displaying
a sharp anomaly $\Delta \chi_3 = \chi_{3} (T_{c}^{-})-\chi_{3}
(T_{c}^{+})$ 
that tracks closely with the
structure of the specific heat \cite{Miyako91,Ramirez92}. 
The non-spinflip ($\Delta J_z =0 $) 
magnetic excitations
seen in both inelastic neutron scattering \cite{Broholm91} and in Raman
measurements \cite{Buhot14,Kung15} also have Ising character, despite
the absence of local moments at those temperatures and pressures.
Finally, quantum oscillations measured deep within the HO region
indicate a quasiparticle g-factor with strong Ising
anisotropy, $g(\theta) \propto \cos \theta$, where $\theta$ is the
angle away from the c-axis \cite{Ohkuni:1999ig,Altarawneh:2011hia}.
This $g(\theta)$ is confirmed by upper critical field
experiments \cite{Altarawneh:2012cy}, that indicate that Ising
quasiparticles pair to form a Pauli-limited superconductor.  In this
paper, we present a \emph{bulk thermodynamic measurement} of the Ising
nature of the hidden order parameter, which shows that this Ising
anisotropy is present not only deep inside the HO,
but at the transition itself; it is  even present
in the order parameter fluctuations above $T_c$.

\figwidth=\columnwidth
\fg{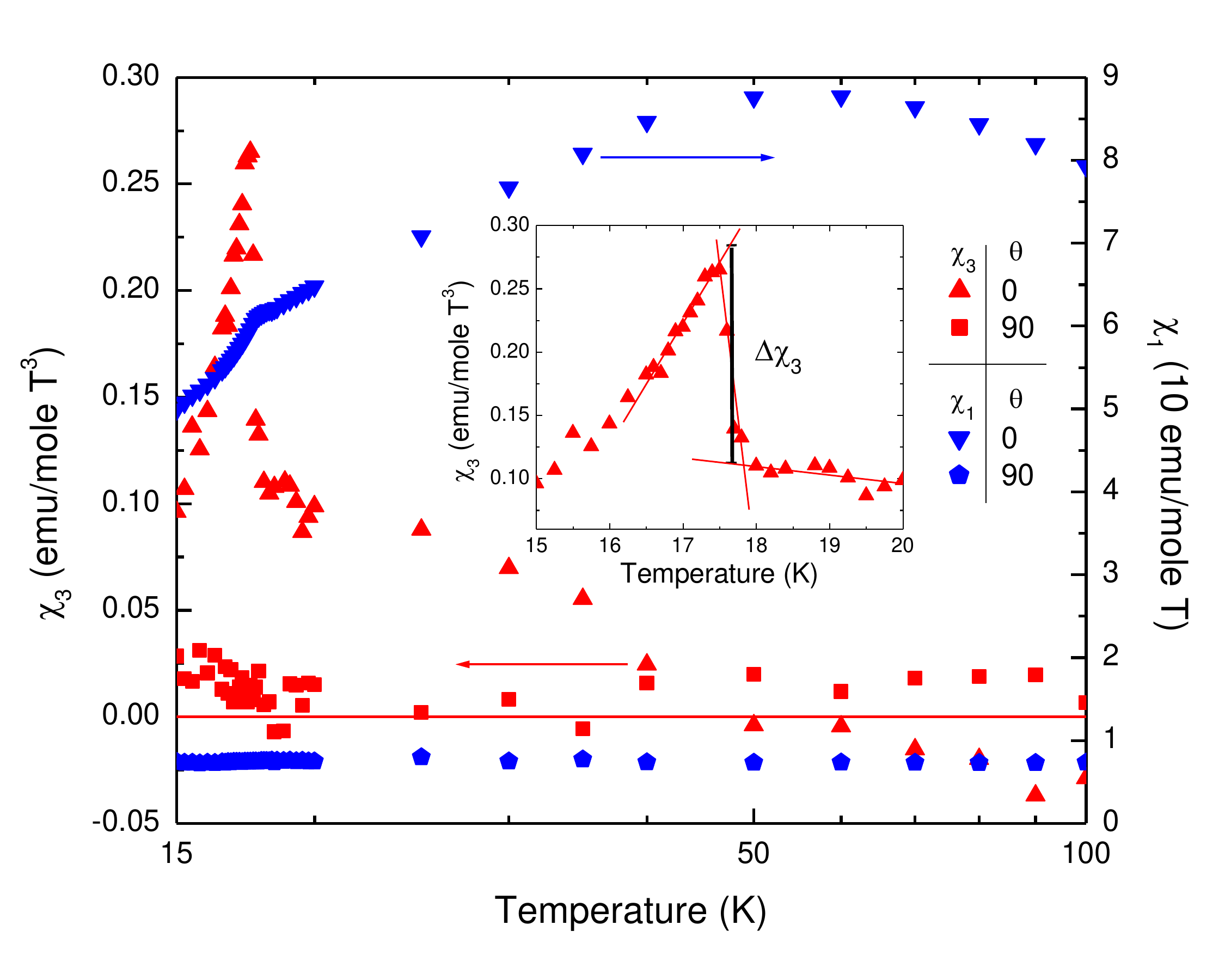}{fig1fin}{Showing linear and nonlinear susceptibility versus
temperature for fields along the c-axis ($\theta =0^{\circ}$) and in the basal
plane ($\theta = 90^{\circ}$).}

As a rank-4 tensor, the nonlinear susceptibility 
$\chi_{3abcd}$,
\begin{equation}
M^a = \chi_{1ab} H^b + \frac{1}{3!}\chi_{3abcd} H^b H^c H^d
\end{equation}
is particularly well-suited to probe symmetry-allowed anisotropies in 
the tetragonal crystal environment (space group $I4/mmm$) of \urs; here 
$M$ and $H$ refer to the magnetization and the applied magnetic field 
respectively, and we use a summation convention for repeated indices. 
In this paper we present an angular survey of the HO 
transition, reporting an extensive series of nonlinear 
susceptibility [$\chi_3 (\theta,\phi$)] measurements. Our results
have important implications for the nature of the quasiparticles
in the HO phase, and we also use $\chi_3(\theta,\phi)$ to probe
the angular anisotropy of short-range order parameter fluctuations at
temperatures above the HO transition.

The general expression for the field-dependent part of the free
energy in a tetragonal crystal at fixed temperature is
\begin{equation}
F = - \chi_1 (\theta) \frac{H^2}{2} - \chi_3 (\theta,\phi) \frac{H^4}{4!} 
\end{equation}
with
\begin{align}
\chi_1 (\theta) & = \chi_1^a + \chi_1^b \cos^2\! \theta \quad \mathrm{and}\cr
\!\!\!\!\!\!\!\!\chi_3 (\theta,\phi) & =  
\chi_3^a 
+  \chi_3^b \cos^2\! \theta 
+ \chi_3^c \cos^4\! \theta 
+ \chi_3^d \sin^4\! \theta \sin^2\! 2\phi
\label{chi3theta}
\end{align}
where $\theta$ and $\phi$ refer to the 
angles away from the c-axis and in the basal plane respectively; details
of this angular decomposition are in the Supplementary Material. 
The anomaly in $\Delta \chi_{3}$
is a known signature of HO \cite{Ramirez92}.
Because there is no Van Vleck contribution to the anomaly $\Delta \chi_3$, 
it is a direct thermodynamic probe of the g-factor at 
the HO transition.  
A key question
is whether the anisotropic g-factor found in quantum oscillations persists to higher temperatures
in the hidden order phase.  Consistency with the low-temperature $g(\theta) \propto \cos \theta$
results requires a large change in $\chi_3^c$, $\Delta \chi_3^c$, and negligible $\Delta \chi_3^a$ and
$\Delta \chi_3^b$. 
 



The URu$_2$Si$_2$ crystal used in this study is of dimension
4mm$\times$2.5mm$\times$2mm and has been previously described
\cite{Ramirez92,Ramirez91}. A recent measurment of $C(T)$ as well 
as the $\chi_1$ and $\chi_3$ measurements reported here show no change in these properties over time
\cite{Ramirez92}. The narrow width of the specific heat transition, $\Delta
T_{HO}= 0.35K$ is consistent with high quality samples of comparable
dimensions \cite{Matsuda11,Kambe13,Niklowitz15}.  
Additionally, the single
superconducting transition indicates a single phase \cite{Ramirez91}
confirming the high quality of the sample.
Measurements of the magnetization, $M$, were performed in a commercial
superconducting quantum interference device (SQUID) magnetometer, as a
function of temperature ($T$), magnetic field ($H$), and angle ($\theta$) between the
sample's c-axis and H.  The variation in angle was achieved with a set
of sample mounts machined from Stycast 1266 epoxy.  The linear and
leading nonlinear ($\chi_3$) susceptibilities were determined as in
\cite{Ramirez92}. Multiple measurements ($\sim$1800 M(H) scans) were performed with 
sufficient resolution in $H$, $T$ and $\theta$ to resolve
the angular dependence of the $\chi_3$ discontinuity at $T_{c}$. Values for $\Delta\chi_3$ were 
obtained at every $\theta$ using a straight-line 
construction assuming a mean-field jump at $T_c$.

Figure 1 shows $\chi_1(T)$
and $\chi_3(T)$ as a function of temperature at $\theta = 0^{\circ}$
and $90^{\circ}$, 
data that
agree well with previous reports
\cite{Ramirez92}.  
We note that the nonlinear susceptibility displays a sharp
anomaly at the HO transition, whereas $\chi_{1} (T)$ displays a
corresponding discontinuity in its gradient $d\chi_1(T)/dT$; both $\chi$
and $\chi_3$ are significantly larger for $\theta = 0^{\circ}$ (c-axis)
than for $\theta = 90^{\circ}$ (ab plane). 

In Figure 2 we show the angular
dependence of $\Delta\chi_3$ and of $\chi_1$ just above the HO transition.  
The linear susceptibility displayed in figure 2 is characterized
by the form
\begin{equation}
\chi_1(\theta,T) = \chi_1^{(0)} + \chi_1^{Ising} (T) \cos^2 \theta,
\end{equation}
where
the isotropic component $\chi_1^{(0)}$ of the susceptibility
displays no discernable temperature dependence. 
The temperature-dependent Ising component, $\chi_1^{Ising}$ displays 
a discontinuity $\frac{d\chi_1^{Ising}}{dT}$ at the HO 
transition. Whereas $\chi_1 (\theta)$ varies as $\cos^2 \theta$
at $T = 18 K$, in Fig. 2 the sharp jump in $\chi_3$ at the transition, 
$\Delta \chi_3$ has a distinctive $\cos^4\theta$ dependence 
\begin{equation}
\Delta\chi_3 (\theta,\phi) = \Delta \chi_3^{Ising} \cos^4 \theta
\end{equation}
without any constant (Van Vleck) terms; this then indicates that
$\Delta \chi_3^c >> \Delta \chi_3^b, \Delta \chi_3^c$ 
in (4), consistent with the low-temperature $g(\theta)$ measurements.
We note that, 
within experimental resolution, no 
$\chi_{3}^{d}$ component was observed in the measurements, either above or at
the transition.

\fg{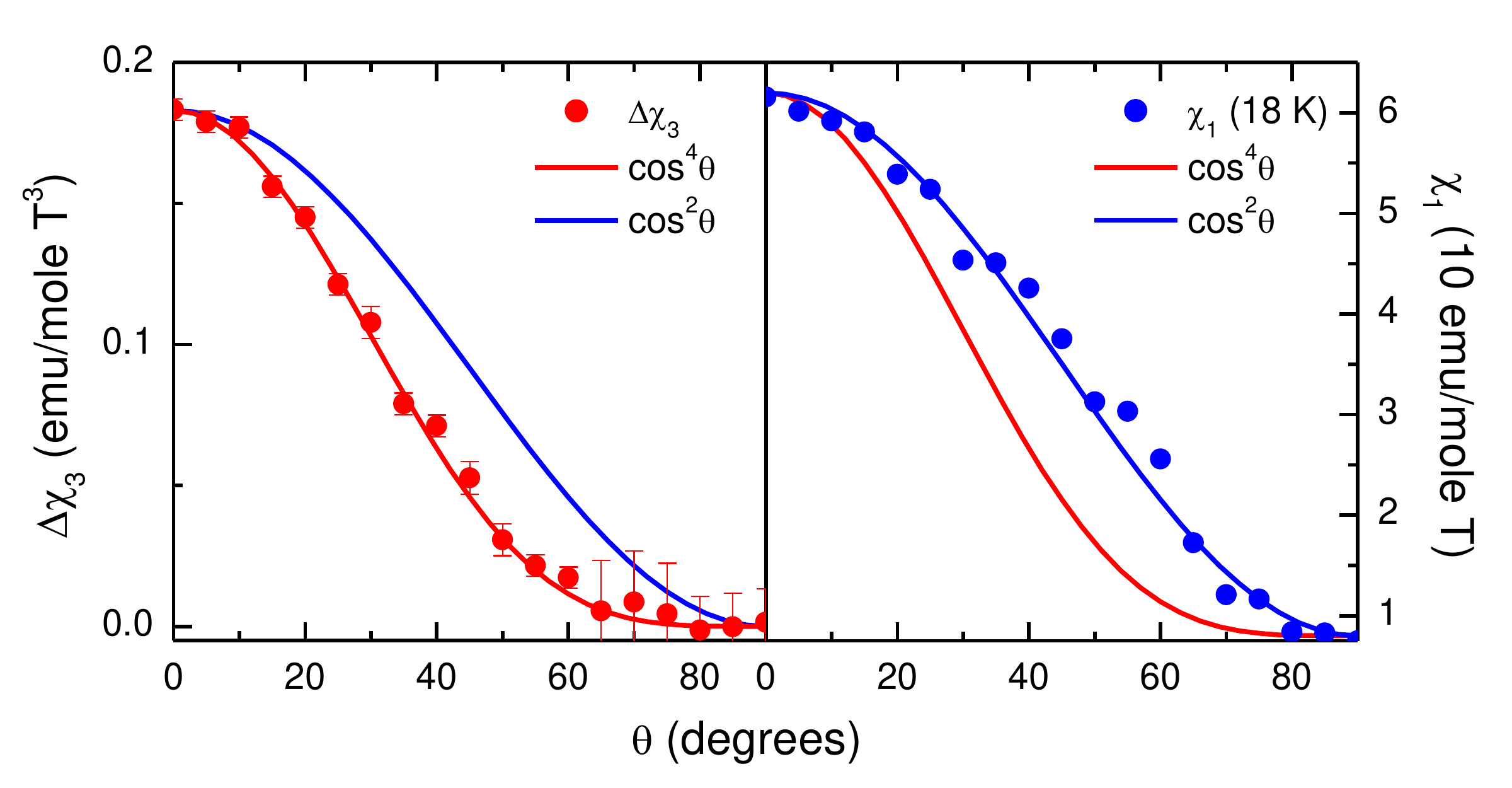}{fig2fin}{Angular dependence of 
(a) the jump $\Delta \chi_{3}$ in
the nonlinear susceptibility at the hidden order transition and 
(b) the magnetic susceptibility just above the hidden order transition
at 18K. }

In Figure 3 we compare the angular dependences of $\Delta\chi_3$ 
with $\chi_3(18 K)$, $\chi_3(30 K)$ and $\chi_3(100 K)$, just above, moderately above and well above $T_{c}$. 
At 18 K and 30 K, $\chi_3$ follows $\cos^4\theta$, similar
to $\Delta\chi_3$.
At 100 K, the positive contribution to $\chi_3$ 
associated with the HO transition has completely vanished,
leaving a negative response presumably associated with
single-ion dipolar physics; the signal is too small to resolve the anisotropy.
At $T = 18 K$,
$\chi_3$ is about 1.6 times smaller 
than $\Delta \chi_3$, (c.f. Fig. 1), 
and is well described by the form 
\begin{equation}
\chi_3(\theta,T) = \chi_3^{(0)} + \chi_3^{Ising} (T) \cos^4 \theta,
\end{equation}
where
the isotropic component $\chi_3^{(0)}$ is essentially temperature-independent. 
A $\cos^4\theta$ dependence in $\chi_{3} (T)$ is still
observed at $30K$, and  by comparing the c-axis and basal plane 
measurements, we estimate that 
around 60K, $\chi_{3}^{Ising} (T)$ goes to zero (see Fig. 1). Above 100K, 
the $\cos^4\theta$ dependence is no longer discernable, leading us to infer 
that the 
Ising component of the nonlinear susceptibility vanishes around 60K.

\fg{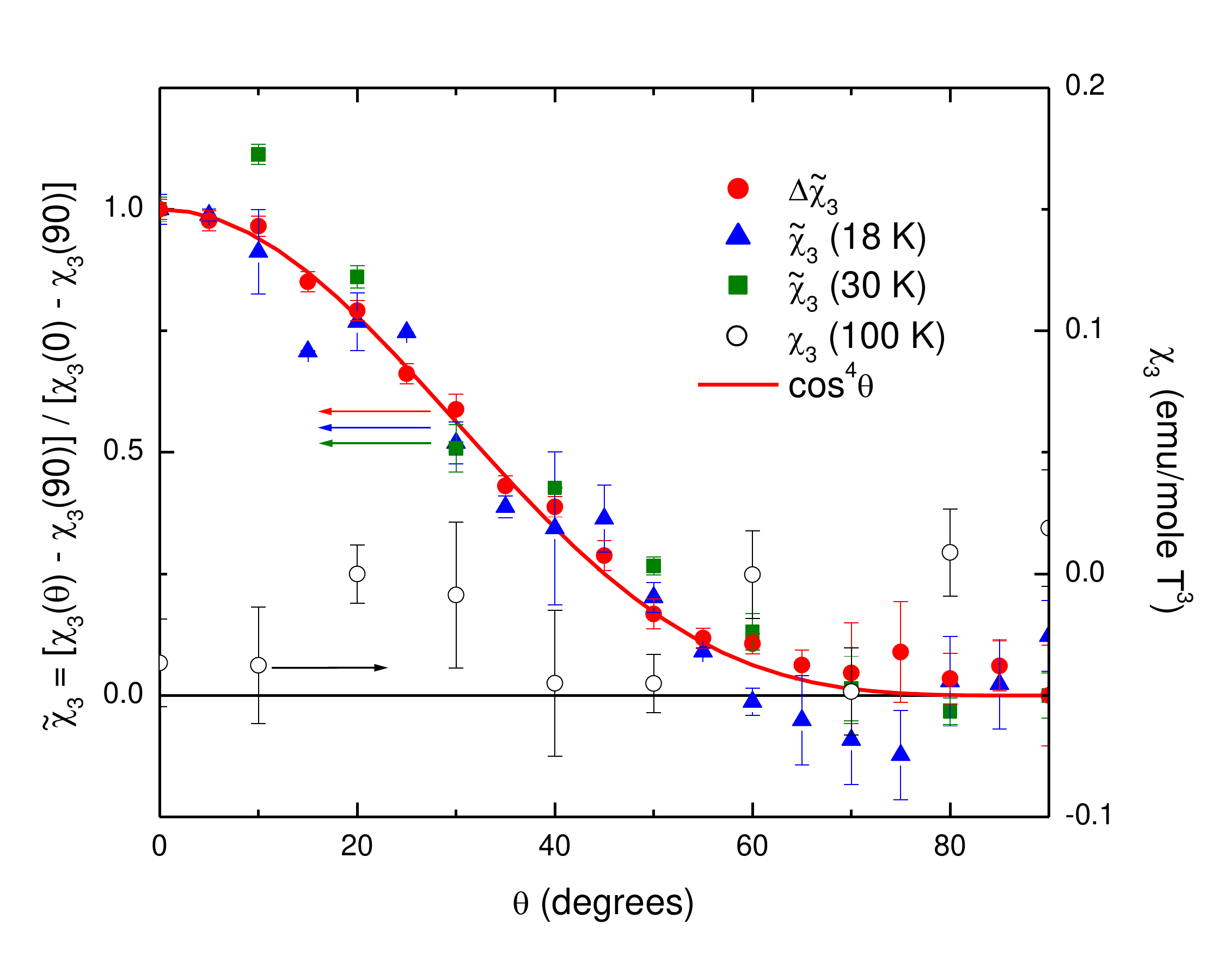}{fig3fin}{Angular dependence of $\Delta \chi_{3}$ and $\chi_{3}$
at three different temperatures, showing the disappearance of the Ising
behavior at high temperatures. }

At the HO transition, our results can be analyzed within a minimal Landau free
energy density of the form 
\begin{equation}\label{landau}
f (T,\psi) = a (T-T_{c} (H) ) 
\psi^{2}
+ \frac{b}{2}\psi^{4}
\end{equation}
where we 
describe a domain of hidden order
by a real order parameter $\psi$ and
\begin{equation}\label{}
T_{c} (H)=
T_{c }-\frac{1}{2}Q_{ab}H_{a}H_{b} +O (H^{4})
\end{equation}
defines the leading field-dependent anisotropy in the transition temperature, where $Q_{ab}$ is a tensor capturing how the order parameter $\psi$ couples to magnetic field; experimental consequences of (\ref{landau}) \cite{Ramirez92} are
discussed in the Supplementary Material.
The quantity 
$\Delta \chi_{ab}= - a  Q_{ab}\psi^{2}$ is the magnetic susceptibility 
associated with the hidden order. By minimizing the free energy with respect to $\psi $, the free energy
below $T_{c} $ is then $f (T) = - {(a^{2}/2b)}[T_{c} (H) -T]^{2}$.
The jump in the linear and nonlinear susceptibilities are then given
by 
\begin{eqnarray}\label{l}
\left(\Delta \frac{d\chi_{1}}{dT}
\right)_{ab} &=& - \frac{a^{2}}{b} 
{Q_{ab}}
\\
(\Delta \chi_{3})_{abcd}&=&  \frac{a^{2}}{b}
{(Q_{ab}Q_{cd}+Q_{ac}Q_{bd}+Q_{ad}Q_{cb})}.
\end{eqnarray}

In order to determine the robustness of the Ising anisotropy, by setting
$Q_{xx}=Q_{yy}=\Phi Q_{zz}$, we codify our results 
in terms of an angle-dependent coupling between the hidden order
parameter $\psi $ and the magnetic field of the form
\begin{equation}\label{}
\Delta  f [\psi ,\theta ] = 
- \frac{a Q_{zz}}{2}\psi^{2}H^{2} (\cos^2\theta+ \Phi \sin^2\theta),
\end{equation}
where $\Phi$ quantifies the fidelity
of the Ising-like behavior, so that $\Phi = 0$ and $\Phi = 1$ 
correspond to Ising and isotropic behavior respectively. The
corresponding jump in the nonlinear susceptibility 
at $T_{c}$ is
\begin{equation}\label{fitting}
\Delta \chi_{3} (\theta ) \propto (\cos^2\theta+ \Phi \sin^2 \theta )^{2}.
\end{equation}  
Our measurements indicate a very small $\Phi = 0.036\pm0.021$, as shown in figure 4 (inset), 
where details of the fitting procedure are given in the Supplementary
Materials. Such a small value of $\Phi$ could be accounted for by an 
angular offset of only one degree, via equation \ref{fitting}. 
X-day diffraction orientation measurements indicate an uncertainty in the c-axis of our sample 
of no more than $\pm3^{\circ}$. In figure 4, one can see that this value provides upper and 
lower bounds to the $\cos^4\theta$-dependence of $\Delta\chi_3$ that bracket the data symmetrically.
Thus, a $\Phi$ value of 0.036 is well below the total uncertainty of the measurement.  
To reduce the uncertainty in $\Phi$ even further would require angular accuracy of well below $1^{\circ}$,
which is beyond the capability of the present apparatus. 
Thus, the obtained $\Phi = 0.036$ is consistent also with $\Phi = 0$.

\fg{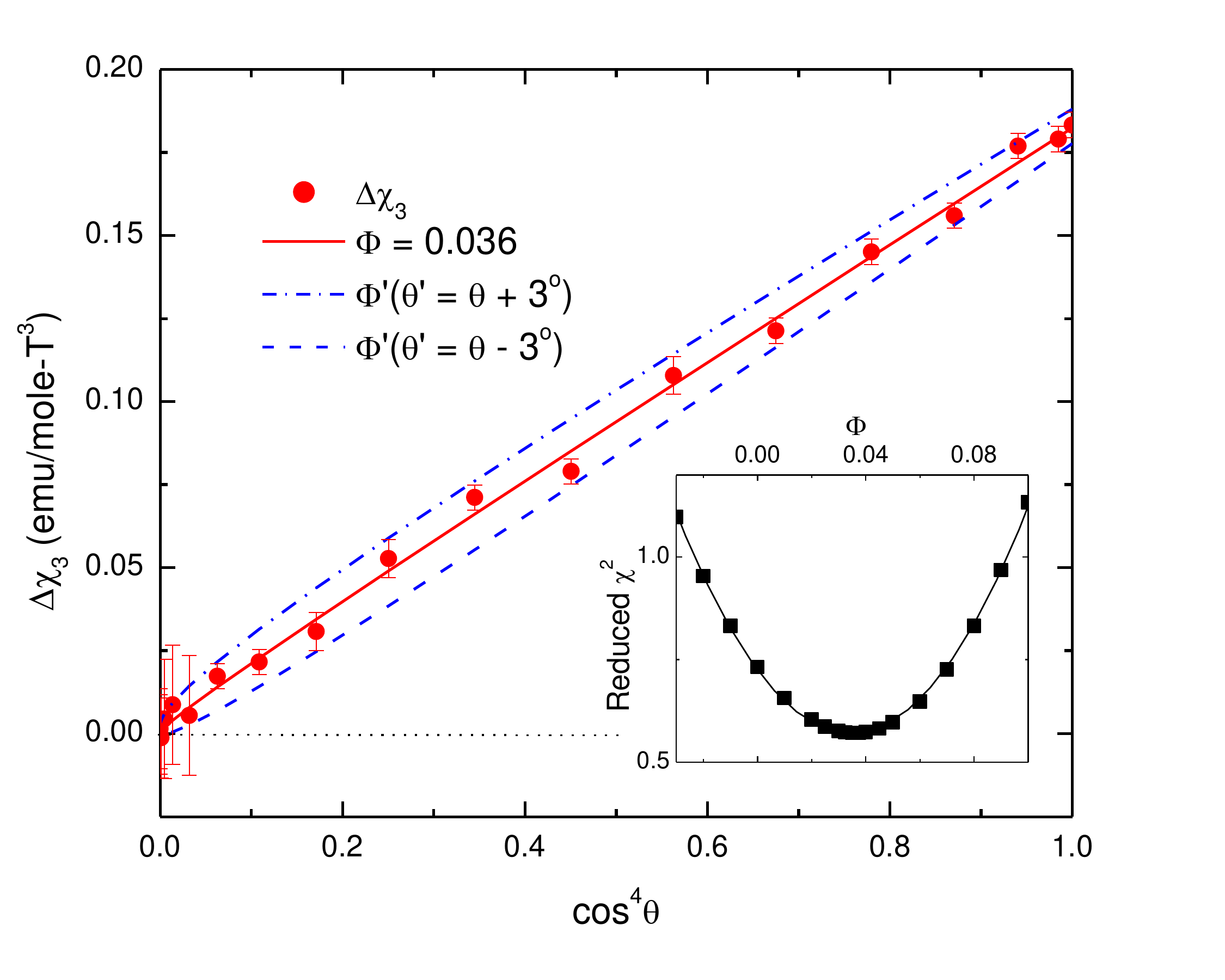}{fig4fin}{$\Delta \chi_{3}$ as a function of $\cos^4\theta$, fit to
equation \ref{fitting} for different values of $\Phi$. Blue dashed lines indicate 
$\Phi$-values assuming an angular offset of $3^{\circ}$ ($\theta' = \theta\pm3^{\circ}$). 
Inset shows the effect of $\Phi$ on the goodness of fit, 
expressed as the reduced $\chi^2$ (see Supplementary
Material).}

We now discuss the implications of these results. At the very simplest
level, our results show that the 
free energy of \urs only depends on the $z$ component of the
magnetic field, i.e. $F[\vec{H}] = F[H_{z}]$. In particular:
\begin{itemize}
\item The coupling of the order parameter
to the magnetic field involves an Ising coupling $F[H,\psi ] = -\frac{1}{2}Q_{zz}\psi^{2}H_{z}^{2}$ coupling. 

\item In the microscopic Hamiltonian, the 
Zeeman coupling of the magnetic field is strongly Ising, with the
field coupling to the z-component of the total angular
momentum $-J_{z}B_{z}$.
\end{itemize}
The second point follows because derivatives of the free energy with
respect to field are equivalent, inside the trace of the partition
function, to the magnetization operator $-\frac{\delta }{\delta
H_{z}}\equiv \hat M_{z}$, so that if the free energy only depends on
$H_{z}$, the partition function $Z = {\rm Tr}e^{- \beta H}$ 
and hence the Hamiltonian  only depends on $\hat M_{z}= g \hat
J_{z}$. 

However, these simple conclusions have  implications for
the microscopic physics. 
On the one hand, we can  link the Ising anisotropy of the microscopic
Hamiltonian to the single-ion properties of 
the $U$ ions in \urs\!\!, where
the Zeeman coupling $-g_{f}\mu_{B}J_{z}B_{z}$ is a sign of
vanishing matrix elements 
$\langle +\vert  J_{\pm }\vert - \rangle =0 $. 
From a single-ion standpoint, an almost perfect Ising anisotropy is a strong
indication of an {\sl integer spin} 
$5f^{2}$ $U^{4+}$ ground-state with $J=4$. 
High-spin Ising configurations of the alternative
$5f^{3}$ $U^{3+}$ ionic configurations 
are ruled out because the coupling of the
local moment to the tetragonal environment mixes 
configurations by adding 
angular momenta in units of $\pm 4\hbar$, for example  $J_{z}= \pm 5/2$
and $\mp 3/2$,  leading almost
inevitably to a non-zero transverse Zeeman coupling when the angular
momentum $J$ is half-integer. 
Although the precise crystal-field configuration of the U ions is still
uncertain \cite{Jeffries:2010,Wray:2015gga,Booth:2016}, both dynamical mean-field
calculations\cite{Haule09} and
high-resolution RIXS measurements \cite{Wray:2015gga} 
confirm the predominantly $5f^{2}$ picture.

Yet a single-ion picture is not enough, for
the sharpness of the specific-heat anomaly, the
sizable entropy and the gapping of two-thirds of the Fermi surface
associated with the hidden order transition\cite{Mydosh:2011wm}
all suggest an underlying
{\sl itinerant} ordering process.  The remarkable feature of our data
is that the jump $\Delta \chi_{3}$ that reflects the itinerant
ordering process exhibits a strong Ising anisotropy. 
This result links  in with the 
observation of
multiple spin zeroes in dHvA measurements, which 
detect the presence of itinerant heavy quasiparticles with an Ising g-factor $g (\theta)=
g_{f}\cos\theta$ at low temperatures.  Our new results suggest that
these same quasiparticles survive all the way up to the hidden order
transition.  In the Landau theory, we can identify the
Ising-like coupling between HO and the magnetic 
field in terms of the squared g-factor $Q_{zz}\psi^{2}\cos^2\theta\propto g
(\theta)^{2}\psi^{2}$.

Reconciling the single-ion and itinerant perspectives,
both supported strongly by experiment, poses a fascinating paradox.
The simplest possibility is that the Ising anisotropy of the
f-electrons is a one-electron effect resulting from a renormalized,
spin-orbit coupled f-band that develops at temperatures well above the
hidden order transition. In this purely itinerant view, the hidden
order is a multipolar density wave that develops within a pre-formed
band of Ising quasiparticles \cite{Rau:2012ig,Ikeda:2012jn}.
Microscopically such quasiparticles are renormalized one-particle
f-orbitals formed from high-spin orbitals with half-integer
$|J_{z}|$. Provided only one $|J_{z}|>1/2$ is involved, the transverse
matrix elements of the angular momentum operator $\langle \pm \vert
J_{\pm }\vert \pm \rangle =0$ identically vanish, leading to a perfect
Ising anisotropy.  Such Ising quasiparticles have 
been observed in strong spin-orbit coupled systems, but only at 
high symmetry points in the Brillouin zone \cite{Saito16}.
Moreover,
in a tetragonal environment, when an electron resonantly
scatters off an f-state, $J_{z}$ is only conserved
$\hbox{mod} (4)$.  Thus a mobile heavy Bloch wave must actively 
exchange $\pm 4 \hbar $ units of angular momentum as it 
propagates through the lattice, leading to Bloch states composed of a mixture of $J_{z}$ states, 
such as
\begin{equation}
\vert \bk \pm \rangle = 
\alpha \vert \bk , \pm 5/2\rangle  + \beta \vert \bk, \mp
3/2\rangle.
\end{equation}
This inevitably gives rise to a finite transverse coupling and a
finite $\Phi$  in the phenomenological Landau theory,
($\Phi \propto |\alpha \beta |^{2}$) that is ruled out by these experiments. 

An alternative is that the itinerant f-quasiparticles carry
{\sl integer} angular momentum, inheriting the Ising anisotropy of 
a localized $5f^{2}$ local moment of the U atoms 
via a phase transition rather than a crossover.   
In this scenario, even though $J_{z}$ is conserved $\hbox{mod}
(4)$, Ising anisotropy is preserved since the up-spin and down-spin
configurations differ by at least two units of angular momentum.
However, this picture requires that 
the half-integer conduction electrons hybridize with the underlying
integer f-states, which can only occur in the presence of a spinorial
or ``hastatic'' order 
parameter \cite{Chandra:2013gv,Flint:2014en,Chandra14,Chandra15}. 
Indeed, the hastatic order scenario predicted
the $\Delta \chi_3\propto \cos^4 \theta$ observed in this
experiment, although theoretical efforts to develop a microscopic theory of hastatic order predicted a small transverse moment that has been shown to be absent in high-precision neutron
scattering experiments \cite{Das:2013em,Metoki:2013ix,Ross:2014do}.  
The vanishing of the anisotropy constant ($\Phi=0$) in our nonlinear susceptibility
measurements combined with the null result reported by neutron scattering
represents a fascinating challenge to our future understanding of
hidden order.

The continuation of the Ising anisotropy well above $T_c$ is also remarkable.  While single-ion physics can give a negative Ising anisotropic $\chi_3$, for an isolated Ising ground state doublet, or a positive, but more isotropic $\chi_3$, if there are several singlets in the temperature range of interest, there is no way to explain the positive Ising anisotropic $\chi_3$ emerging below 60K with single ion physics.  Instead, this response indicates Ising anisotropic order parameter fluctuations extending up to more than three times $T_c$, an extraordinarily large fluctuation regime.

An interesting question raised by our work is whether
bulk nonlinear susceptibility measurements can be used to detect
microscopic broken tetragonal symmetry that has been reported in
torque magnetometry measurements \cite{Okazaki:2010tn}. 
In principle, were the hidden order to possess domains with broken
tetragonal symmetry, inter-domain fluctuations in the basal-plane
susceptibility
would manifest themselves 
through a finite value of $\chi^{d}_{3}$  below $T_{c}$.
The large Ising anisotropy suppresses the precision for in-plane
susceptibility measurement: our current work 
places an upper bound on the 
microscopic symmetry-breaking susceptibility $|\Delta \chi_{xy}|$ 
such that
$|\Delta\chi_{xy}|/\chi_{xx} \le 1$ that is two orders of magnitude 
larger than that measured by 
torque magnetometry on $\mu $m size samples\cite{Okazaki:2010tn}  
(see Supplementary Materials), and thus our negative results are 
not inconsistent with their positive finding.
However improvement in resolution 
in future measurements could 
make it possible to address this issue.

In summary, we have presented a detailed survey of the nonlinear
magnetic susceptibility as a function of angle and temperature in the
hidden order compound \urs\!\!.  These measurements showcase the unique
Ising anisotropy, and imply that it is a key feature
of the hidden order parameter.  While previous quantum oscillation
measurements indicated the presence of Ising quasiparticles, this
Ising anisotropy persists not only to the transition temperature, but all the way up to 60K, putting serious constraints on the theory of hidden order.  It would be quite interesting to examine the nonlinear susceptibility anisotropy in and above the antiferromagnetic phase, which could be done in URu$_{2-x}$Fe$_x$Si$_2$\cite{Kanchanavatee:2011}.

{\sl Acknowledgments:} 
This work was supported by National Science Foundation grants
NSF DGE 1339067 (J. Trinh), NSF DMR-1334428 (P. Chandra), NSF DMR 1309929 
(P. Coleman), NSF DMR 1534741 (A.P. Ramirez) and Ames Laboratory
Royalty Funds and Iowa State University startup funds (R. Flint). The Ames Laboratory is operated for the U.S. Department of Energy by Iowa State University under Contract No. DE-AC02-07CH11358. PC, PC and  
and RF acknowledge the hospitality of the Aspen Center for Physics, 
supported by NSF PHYS-1066293, where early parts of this work 
were discussed. TS acknowledges funding by the 
U.S. Department of Energy, Office of Basic Energy Sciences, Materials Sciences and Engineering Division, under award No. DE-SC0008832. P. Chandra thanks S. Bahramy for a stimulating
discussion on Ising quasiparticles in $MoS_2$.  
%
\newpage
\begin{widetext}
\section{Supplementary material for ``Thermodynamic Measurement of
Angular Anisotropy at the Hidden Order Transition of \urs''}

\maketitle

\section{Angular decomposition of $\chi_{3}$ in a tetragonal
environment}\label{}

The expansion of the Free energy as  a function of magnetic field
strength $H$  can be written
\begin{equation}\label{}
\Delta F = - \frac{1}{2}H_{a}H_{b}\chi^{1}_{ab} 
-
\frac{1}{4!}H_{a}H_{b}H_{c}H_{d}\chi^{3}_{abcd}
\end{equation}
The fourth order coupling to the field can be treated in a fashion 
closely analogous to the elastic free energy of a material, 
$\chi^{3}_{abcd}$ playing the role of the elasticity tensor
$C_{abcd}$, with the additional proviso that $\chi^{3}_{abcd}$ is
a completely symmetric tensor. 
In  a tetragonal crystal with mirror symmetries,  the only non-vanishing
components of the nonlinear susceptibility tensor are 
\begin{eqnarray}\label{l}
\chi_{1111}&=&\chi_{2222},\qquad \cr\chi_{3333}&&\cr
\chi_{1122}&=& \chi_{1212},\cr
\chi_{1133}&= &\chi_{2233}= \chi_{1313} = \chi_{2323}.
\end{eqnarray}
The fourth order contribution to the nonlinear susceptibilty can then
be written as 
\begin{eqnarray}\label{l}
\Delta F_{4} &=& -\frac{1}{4!}
H_{a}H_{b}H_{c}H_{d}\chi^{3}_{abcd} \cr &=& 
-\frac{1}{4!}
\left[ 
\chi_{1111} (H_{x}^{4}+H_{y}^{4})+ \chi_{3333}H_{z}^{4}+ 6
\chi_{1122}H_{x}^{2}H_{y}^{2} 
+ 6\chi_{1133} (H_{x}^{2}+H_{y}^{2})H_{z}^{2}
\right]
\end{eqnarray}
Substituting $(H_{1},H_{2},H_{3})\rightarrow H( \sin \theta \cos \phi,
,\sin \theta \sin \phi, \cos \theta  )$, we obtain
\begin{eqnarray}\label{l}
\Delta F &=& 
- \frac{H^{4}}{4!} \left(
\chi_{1111} (1-\cos^{2}\theta )^{2} (1 - \frac{1}{2}\sin^{2}2\phi ) 
+ \chi_{3333}\cos^4\theta
\right.
\cr &+&\left. \phantom{\int}
\frac{6}{4}\chi_{1122}\sin^4\theta\sin^2 2 \phi 
+ 6 \chi_{1133} (1 - \cos^2\theta)\cos^2\theta
\right)\cr
&=& - \frac{H^{4}}{4!} 
(
\chi_3^{a} + 
\chi _3^b \cos^2\theta +
 \chi _3^c \cos^4\theta 
+
\chi_{3}^{d}sin^4\theta  \sin^{2 }2\phi 
)
\end{eqnarray}
where
\begin{eqnarray}\label{l}
\chi_{3}^{a}&=&\chi_{1111}, \cr
\chi_{3}^{b}&=&6 \chi _{1133}
-2 \chi _{1111}\cr
\chi_{3}^{c}&=&\chi _{1111}+\chi _{3333}-6\chi_{1133}
\cr
\chi_{3}^{d}&=&\frac{1}{2} \left(-\chi _{1111}+3\chi
   _{1122}\right).
\end{eqnarray}

\section{Landau theory for the minimal free energy}\label{}

In this section, we further document the experimental consequences of our minimal Landau free energy, as has already been explored in Ramirez \emph{et al}\cite{Ramirez:1992ws}.  If we take the free energy,
\begin{equation}
f[T,\psi] = a (T-T_c+\frac{1}{2}Q_{zz}H_z^2)\psi^2 + \frac{b}{2}\psi^4,
\end{equation}
and solve for $\psi^2(T,H_z) = a^2(T_c-\frac{1}{2}Q_{zz}H_z^2-T)^2/b$, we find the free energy below $T_c$,
\begin{equation}
f[T] = -\frac{a^2}{2b}(T_c-\frac{1}{2}Q_{zz}H_z^2-T)^2.
\end{equation}
We can then take the appropriate derivatives to obtain the jumps in the specific heat, $d\chi_1/dT$ and $\chi_3$ at $T_c$ in zero field,
\begin{equation}
\frac{\Delta C_V}{T_c}=\frac{a^2}{b}; \quad \Delta \frac{d\chi_1}{dT} = \frac{a^2 Q_{zz}}{b}; \quad \Delta \chi_3 = \frac{3 a^2 Q_{zz}^2}{b}.
\end{equation}
These jumps obey the thermodynamic relation, 
\begin{equation}
\frac{\Delta C_V}{T}\Delta \chi_3 = 12\left(\frac{d\chi_1}{dT}\right)^2.
\end{equation}
As these quantities have been measured for hidden order ($\frac{\Delta C_V}{T_{HO}} = 300$ mJ/mol K$^2$\cite{Palstra:1985wa} and $\Delta \chi_3 = 1.8$ emu/mol T$^3$, we can estimate,
\begin{equation}
Q_{zz} = \sqrt{\frac{\Delta \chi_3 T_{HO}}{3\Delta C_V}} = .04 \mathrm{K/T}^2.
\end{equation}
If one neglects higher order corrections to $T_c(H_z) = T_c -\frac{1}{2}Q_{zz}H_z^2+O(H_z^4)$, this value of $Q_{zz}$ means that $T_{c}$ would vanish around 28T, which is roughly consistent with the experimental value\cite{Kim:03}, even though higher order terms will certainly be important at those high fields.

\section{Angular fitting procedure for $\Delta\chi_{3} (\theta )$}\label{}

We assumed a fit for $\Delta \chi_3$ of the form 
\begin{equation}\label{fitting}
\Delta \chi_{3} (\theta ) \propto (\cos^2\theta+ \Phi \sin^2 \theta )^{2}.
\end{equation}  
and determined $\Phi$ by optimizing the goodness of fit, or
by minimizing the reduced $\chi^2$ of the fit, which is
defined as the residual sum of squares 
weighted against the variance (square of the error) at each point, 
divided by the number of data points less the number of fit parameters,  
\begin{equation}
\chi_{red}^2 = \sum_{i=1}^{n}\frac{\sigma^{-2}(\Delta\chi_{3 meas}(\theta_i)-\Delta\chi_{3 fit}(\theta_i))^2}{n - m}.
\end{equation} 
Here $\sigma^2$ is the variance, n is the number of data points and m is the number 
of fit parameters, which in this case is 1 since $\Phi$ is manually adjusted. Weighting against
the variance allows us to incorporate the error in each $\Delta\chi_3$ measurement
in our fit, while dividing by $n-m$ normalizes for the degrees of freedom. 

The reduced $\chi^2$ was calculated for several values of $\Phi$, and the 
results are shown in the inset of Fig. 4 (main text), 
where the minimum as determined by a quadratic fit is $\Phi = 0.036\pm0.021$.  
As mentioned in the main text, an angular offset of one degree ($\theta' = \theta + 1^{\circ}$) 
is sufficient to account for this value of $\Phi$ when fitting to equation \ref{fitting}. 

The intent behind using the reduced $\chi^2$ is not to determine the goodness of fit relative to 
other models, but to extract an estimate for $\Phi$ and its error given the model that we propose.
Thus, it is the robustness of the \textit{minimum} of the reduced $\chi^2$ that we require, not
its absolute value. To ensure that our value of $\Phi$ is insensitive to random error, we have 
examined the effect of masking various points. The removal of any one point from the data set
resulted in a change in $\Phi$ of $\pm 0.0013$ on average, less than our uncertainty in $\Phi$
of 0.021.

Addtitionally, we compared our results with a similar analysis using other 
measures for goodness of fit (e.g., $R^2$) and found $\Phi$-values within $\pm{0.0006}$
of 0.036, which is again within our error estimate. 

\section{Error bounds on the in-plane anisotropy of $\chi_{3}$
}\label{}
Here we compute the bounds that our nonlinear susceptibility
measurement place on the magnitutude of the in-plane tetragonal
symmetry breaking. 
For a single domain, broken tetragonal symmetry breaking manifests
itself through the development of a finite off-diagonal component of
the magnetic susceptibility, denoted by 
\begin{equation}\label{C1}
\chi_{xy}^{D}\sim \left(\frac{V_{D}}{v_{c}} \right) \frac{\langle m_{x}m_{y}\rangle }{T},
\end{equation}
where $V_{D}$ is the volume of the domain, $v_{c}$ is the volume of a
unit cell and $m_{a}= M_{a}/N_{cells}$ is the magnetization per cell.
Now the 
bulk off-diagonal magnetic susceptibility involves an average over
many
different domains, which is zero:
\begin{equation}\label{}
\overline{\chi_{xy}}=0,
\end{equation}
where the overbar denotes a domain average. 
However, the domain fluctuations in the susceptibility remain finite,
given by 
\begin{equation}\label{}
\overline{(\Delta \chi_{xy})^{2}}= \left(\frac{V}{V_{D}} \right) (\chi_{xy}^{D})^{2},
\end{equation}
where $V$ is the total volume of the sample, 
and it is these fluctuations that give rise to a component of
$\chi_{3}^{d}$. 
Now the change in the bulk basal-plane nonlinear susceptibility (i.e
in the ab-plane, perpendicular to the c axis) in the hidden
order phase 
is then given by 
\begin{equation}\label{}
\overline{\Delta \chi^{3}_{\perp }}
\sim - \left(\frac{V}{v_{c}}
\right)\frac{\overline{\langle m_{x}m_{y}\rangle^{2}}}{T^{3}}.
\end{equation}
Substituting in (\ref{C1}), we then obtain
\begin{equation}\label{C2}
\overline{\Delta \chi^{3}_{\perp }}\sim 
- \left(\frac{Vv_{c}}{V_{D}^{2}}
\right)\frac{\left(\chi_{xy}^{D} \right)^{2}}{T}= 
- \left(\frac{v_{c}}{V_{D}}
\right)\frac{\overline{(\Delta \chi_{xy})^{2}}
}{T}
\end{equation}
Thus the inter-domain fluctuations in the symmetry-breaking component 
of the nonlinear susceptibilty are expected to generate a
contribution to $\Delta \chi^{3}_{xyxy}$.

To set bounds on this, we compare the anomalous basal-plane component of the
nonlinear susceptibility with the nonlinear susceptibility along the
z-axis, given by 
\begin{equation}\label{}
\chi^{3}_{zzzz}= \left(\frac{V}{v_{c}} \right) \frac{\langle m_{z}^{4}\rangle }{T^{3}}
\end{equation}
Now the susceptibility in the z-direction of a single domain is given
by 
\[
\chi_{zz}^{D} = \frac{V_{D}}{v_{c}} \frac{\langle m_{z}^{2}\rangle}{T}
\]
so that we can write
\begin{equation}\label{C3}
\chi^{3}_{zzzz}\sim\left(\frac{V}{v_{c}} \right) \left[\left(\frac{\chi_{zz}^{D}v_{c}}{V_{D}} \right) \right]^{2}\frac{1}{T}
\end{equation}
Taking the ratio of (\ref{C2}) and (\ref{C3}) we obtain 
\begin{equation*}
\frac{
\overline{\Delta \chi^{3}_{\perp }}
}{\chi^{3}_{zzzz}} 
= -\left(\frac{\chi_{xy}^{D}}{\chi^{D}_{zz}}
\right)^{2}
= -\left(\frac{\chi_{xy}^{D}}{\chi^{D}_{xx}}
\right)^{2}
\left(\frac{\chi_{xx}^{D}}{\chi^{D}_{zz}}
\right)^{2}
= -\left(\frac{\chi_{xy}^{D}}{\chi^{D}_{xx}}
\right)
^{2}
\left(\frac{\chi_{xx}}{\chi_{zz}}
\right)
^{2}
\end{equation*}
where we have removed the superscript ``$D$'' in the ratio between
basal-plane and c-axis susceptibilities in the last term.
We thus see that the magnitude of the anomalous basal plane nonlinear
susceptibility is substantially reduced by the squared ratio of the bulk 
basal-plane and c-axis susceptibilities.

We can rearrange this equation to set bounds on the in-plane
tetragonality as follows:
\begin{equation}\label{}
\left|\frac{\chi_{xy}^{D}}{\chi^{D}_{xx}}
\right| \leq  
\left(\frac{\chi_{zz}}{\chi_{xx}}
\right)
\sqrt{
\left\vert 
\frac{
\overline{\Delta \chi^{3}_{\perp }}
}{\chi^{3}_{zzzz}} 
\right\vert 
}
\end{equation}
Putting in numbers, the anisotropy in the linear 
susceptibility is at least five,  
\begin{equation}\label{}
\left(\frac{\chi_{zz}}{\chi_{xx}}
\right) > 5
\end{equation}
while 
the error bounds on the measurement of the in-plane nonlinear
susceptibility are given by 
\begin{equation}\label{}
\left\vert 
\frac{
\overline{\Delta \chi^{3}_{\perp }}
}{\chi^{3}_{zzzz}} 
\right\vert  \leq 0.14
\end{equation}
so that 
\begin{equation}\label{}
\left|\frac{\chi_{xy}^{D}}{\chi^{D}_{xx}}
\right| \leq  5 \times \sqrt{0.14}\sim 1.9
\end{equation}
which sets a bound which is two orders of magnitude larger than the
anisotropy measured by torque magnetometry in micron-sized tiny
samples. 

Thus there is no inconsistency between our nonlinear
susceptibility measurement and previous torque magnetometry
measurements. We also see that an order of magnitude improvement in
the nonlinear susceptibility measurements would make it possible to
observe the in-plane anisotropy using a bulk probe. 
\end{widetext}

\end{document}